\documentclass[aps,prl,preprint,superscriptaddress]{revtex4}

\pdfoutput=1
\usepackage {graphicx}
\usepackage{amsmath}
\usepackage{wrapfig}
\usepackage{epstopdf}
\usepackage{color}

\begin{document}

\title{Theory and simulations for crowding-induced changes in stability of proteins with applications to $\lambda$ repressor}
\author{Natalia A. Denesyuk}
\affiliation{Biophysics Program, 
Institute for Physical Science and Technology, University of Maryland, College Park, MD 20742}
\author{D. Thirumalai}
\affiliation{Department of Chemistry\\
University of Texas at Austin, Austin, TX 78731
}
\date{\today}

\begin{abstract} 
Experiments and theories have shown that when steric interactions between crowding particles and proteins are dominant, which give rise to Asakura-Oosawa depletion forces, then the stabilities of the proteins increase compared to the infinite dilution case. We show using theoretical arguments that  the crowder volume fraction ($\Phi_C$) dependent increase in the melting temperature of globular proteins, $\Delta T_m(\Phi_C) \approx \Phi_C^{\alpha}$ where $\alpha = \frac{1}{(3 \nu_{eff} - 1)}$. The effective Flory exponent, $\nu_{eff}$, relates the radius of gyration in the unfolded state to the  number of amino acid residues in the protein.  We derive  the bound 1.25 $\le \alpha \le$ 2.0. The theoretical predictions are confirmed using molecular simulations of $\lambda$ repressor in the presence of spherical crowding particles. Analyses of previous simulations and experiments confirm the predicted theoretical bound for $\alpha$. We show that the non-specific attractions between crowding particles and amino acid residues have to be substantial to fully negate the enhanced protein stabilities due to intra protein attractive Asakura-Oosawa (AO) depletion potential.   Using the findings, we provide an alternate explanation for the very modest (often less than 0.5 Kcal/mol) destabilization in certain proteins in the cellular milieu.  Cellular environment is polydisperse containing large and small crowding agents. AO arguments suggest that proteins would be localized between large (sizes exceeding that of the proteins) crowders, which are predicted to have negligible effect on stability. {\it In vitro} experiments containing mixtures of crowding particles could validate or invalidate the predictions.
\end{abstract} 
\maketitle

\section*{\large{Introduction}}

The crowded cellular environment could profoundly influence many aspects of interest in biophysics, such as, the folding of proteins \cite{Minton05BJ,Elcock10PlosCompBiol}, RNA \cite{Pincus08JACS,Kilburn10JACS,Denesyuk11JACS,Cheung13COSB,Jeon16SoftMatter}, the formation of oligomers \cite{OBrien11JPCL}, binding of intrinsically disordered proteins \cite{Zosel20PNAS}.   This realization has lead to great efforts to understand  crowding-induced folding of proteins \cite{Minton05BJ,Cheung05PNAS,Dhar10PNAS}, and subsequently RNA \cite{Pincus08JACS,Kilburn10JACS,Denesyuk11JACS,Denesyuk13BioPhysRev,Tyrrell13Biochem,Tyrrell15Biochem,Strulson14RNA,Leamy16QRB}. In most of the {\it in vitro}  studies the effects of macromolecular crowding is mimicked by monodisperse particles, which are not only well controlled model systems but also can be considered as an idealization of the cellular environment.  These studies have shown that the crowding agents enhance the stability, relative to what is found under infinite dilution conditions. The explanation for the increased stability can be traced to a remarkable paper in 1954 by Asakura and Oosawa (AO) \cite{Asakura54JCP} and subsequently in a study published four years later \cite{Asakura58JPolySci}.  They showed that the crowding particles (modeled as hard objects) induces an effective short-range attraction between (depletion potential). This concept when applied to folding of globular proteins suggests that  (predominantly) the entropy of the unfolded state should decrease.  Consequently, the crowding should enhance the stability of the folded protein. 

In contrast to the ESM, in several experimental studies it has been argued that weak soft-interactions (also referred to as chemical interactions), presumably between crowding particles and the residues in the polypeptide chains negate the stabilizing effects due to ever present hard core interactions \cite{Drishti19ChemRev,Sarkar13PNAS,Sapir15CurrOpinColl,Sapir14JPCL,Danielsson15PNAS}. Many of these studies imply that this behavior is universal in the sense destabilization due to crowding is the norm, especially under {\it in vivo} conditions.   Thus, there is a need to theoretically describe folding in a crowded environment, which is complicated because the interplay of a  number of factors such as variations in the sizes, shapes, volume fraction ($\Phi_C$) of the crowding agents, and  non-specific attractive interactions between the crowding particles and the proteins determines the fate of proteins. As a result crowding could increase, decrease, or leave unaltered the stabilities of proteins relative to their values when $\Phi_C = 0$ \cite{Cheung05PNAS,Miklos11JACS,Ghosh10BJ,Ebbinghaus11JPCL,Kang15PRL}.  

Regardless of the complexity of the interactions governing the fate of polypeptide chains in a crowded environment, to a first approximation, the most important effect 
is the exclusion of the volume occupied by the macromolecules to the protein of interest.   If excluded volume interactions between the crowding particles and the proteins 
dominate then  the stability of proteins is enhanced compared to $\Phi_C = 0$ \cite{Minton05BJ,Cheung05PNAS}. As noted earlier this is die to the ESM, which takes into account the greater crowding-induced suppression of the conformational fluctuations 
of the unfolded state relative to those in the native folded state, as Minton showed in an important study~\cite{Minton05BJ}.  

Inspired in part by an experimental study \cite{Gai11JCP}, here we provide theoretical arguments and simulations to quantify the dependence of enhanced stabilities of proteins, expressed in terms of the $\Phi_C$-dependence of the melting temperatures, $T_m(\Phi_C)$, in the presence of spherical crowding agents.  Although the use of spherical crowding agents is a caricature of the cellular environment and perhaps even model agents used in {\it in vitro} studies, a quantitative treatment of the simpler system is a useful first step towards a more realistic description of the heterogeneous milieu containing polydisperse crowding agents with a variety of shapes. We show that the increase in $T_m(\Phi_C)$, relative to the value in the absence of crowding, increases as $\Phi_c^{\alpha}$ where the value of the effective exponent, $\alpha$, is determined by the characteristics of the unfolded states {\it in the absence of crowding agents} ($\Phi_C = 0$). The theoretical arguments lead to an upper and lower bound for $\alpha$ but the precise requires a fit to the experimental data. We validate the predictions using coarse-grained simulations \cite{Hyeon11NatComm} of $\lambda_{6-85}$ repressor in the presence of spherical crowding particles.  The values of $\alpha$ extracted from simulations and experiments are consistent with the predicted bounds. We also examine the plausible effects of the non-specific soft attractions between the crowding particles and the polypeptide chain in order to understand the very modest (less than $k_BT$) destabilization found in the presence of cell lysates.  We argue that the thermodynamic stability of proteins even in complex environment is largely determined by excluded volume interactions. However, as predicted theoretically, the effects of steric interactions on the stability could be negligible (practically no effect), and is determined by the ratio of the size of the unfolded state of the protein when $\Phi_C = 0$ to the size of largest crowders in a soup containing mixture of crowding agents. 

\section*{Theory}

{\bf Melting temperature increases with $\Phi_C$:} The physical arguments hinge on the observation that when $\Phi_C \ne 0$ the polypeptide chain would 
prefer to be localized in a region that is devoid of hard sphere-like crowding particles. On general theoretical grounds \cite{Edwards88JCP,Thirum88PRA,vandershoot98Macro}, it can be shown that the probability of finding a void that is large enough to accommodate a polypeptide chain with radius of gyration $R_g$ decreases exponentially as $\Phi_C$ increases. However, the fluctuations in density of the crowding particles would create a void whose optimal size $D$ (Fig. 1) is likely to be spherical \cite{Honeycutt89JCP}. In this sense the effects of crowding can be approximately mimicked by confining the polypeptide chain to a spherical cavity of size $D$ \cite{Cheung05PNAS}. It is straightforward to  show that 
\begin{equation}
D = R_c\Phi_C^{-{\frac{1}{3}}}
\end{equation}
where $R_c$ is the radius of the spherical crowding particle.  Assuming that the mapping between crowding and confinement is reliable, as appears to be the case based on 
simulations \cite{Cheung05PNAS}, we can estimate the decrease in free energy of the unfolded state due to localization of the polypeptide chain in a spherical cavity as, 
$\Delta F_U(\Phi_c) \approx k_BT (R_g/D)^{1/{\nu}}$, where $\nu$ is the Flory exponent describing $R_g \approx a N^{\nu}$, $k_B$ and $T$ are the Boltzmann's constant and temperature respectively, and $N$ is the number of amino acid residues, and $a$ is  approximately the distance between $C_{\alpha}$ atoms. In using this estimate for $\Delta F_U$, we assumed that the unfolded state can be treated as a self-avoiding walk. A more refined treatment which takes into account intra polypeptide interactions \cite{GrosbergBook,Thirumalai03PNAS}, always present even in high denaturant concentration, shows that
\begin{equation}
\Delta F_U = k_BT (R_g/D)^{\frac{3}{(3 \nu -1)}}.
\end{equation} 
Assuming that the folded state is not significantly affected by the crowding particles, we obtain from Eqs. (1 and 2) the following equation for $\Delta T_m(\Phi_C) = T_m(\Phi_C) - T_m(0)$,
\begin{equation}
\Delta T_m(\Phi_C) \approx \left(\frac{\Phi_C}{R_C^{3}}\right)^{\alpha}
\label{Tm}
\end{equation}
where $\alpha = \frac{1}{(3 \nu_{eff} -1)}$. Because of both finite-size of proteins as well as the presence of intra polypeptide attractive interactions 
(at least between hydrophobic residues) $\nu$ in Eq. (3) should be treated as an effective exponent ($\nu_{eff}$), and could depend on the protein of interest.  However, the 
upper bound for $\nu_{eff}$ is $\approx$ 0.6. The lower bound, at which the folding and collapse temperatures are nearly coincident for single domain 
proteins \cite{Camacho93PNAS}, is $\approx$ 0.5.  Thus, the lower and upper bounds on $\alpha = \frac{1}{(3 \nu_{eff} -1)}$ are 1.25 and 2, respectively. If $\alpha < $ 2, it implies that there is residual structure in the unfolded state. The $\alpha$-exponent in Eq. (3) is determined by the dimensions of proteins in the unfolded 
state, and should be valid for {\it any} protein provided only interactions between crowders and polypeptide chain are relevant. 

{\bf Stability changes as a function of $\Phi_C$ and $R_C$:} In order provide a molecular picture of the stability changes of the protein, which can be computed in simulations, we consider a polymer chain with radius of gyration $R_{\rm g}$ in a  suspension of crowders with density 
$\rho_{\rm C}$. The volume fraction of the suspension $\Phi_{\rm C}$ scales as $\rho_{\rm C}R_{\rm C}^3$. 
When the polymer comes in contact with a crowding particle, the polymer segments become depleted (AO effect)
from the particle surface, resulting in an increase in the polymer free energy. If the crowding particles are much 
smaller than the polymer, $R_{\rm C}\ll R_{\rm g}$, the free energy increase per particle is given by \cite{Odijk00PhysicaA}
\begin{equation}
\frac{\Delta F_1}{k_{\rm B}T}\sim \frac{R_{\rm C}}{R_{\rm g}}
\label{Odijk}
\end{equation}
for ideal polymers,
and by
\begin{equation}
\frac{\Delta F_1}{k_{\rm B}T}\sim \left(\frac{R_{\rm C}}{R_{\rm g}}\right)^{4/3}
\label{deGennes}
\end{equation}
for polymers in good solvent \cite{deGennesbook}. We can rewrite both the results as
\begin{equation}
\frac{\Delta F_1}{k_{\rm B}T}\sim \left(\frac{R_{\rm C}}{R_{\rm g}}\right)^{3-1/\nu},
\label{combo}
\end{equation}
where $\nu$ is the Flory exponent, $R_{\rm g}\sim aN^{\nu}$, $N$ is the number of segments in the polymer and $a$ is the segment size. 

Suppose  that the crowders, which have been expelled from the polymer interior, stay close to the polymer surface
(this assumption is justified below). Then, the number of crowders in contact with the polymer, $n$, can be 
estimated as
\begin{equation}
n\sim\rho_{\rm C}R_{\rm g}^3\sim\Phi_{\rm C}\left(\frac{R_{\rm g}}{R_{\rm C}}\right)^3,
\label{n1}
\end{equation}
so that the total free energy loss due to crowding, $\Delta F$, is 
\begin{equation}
\frac{\Delta F}{k_{\rm B}T} = n \frac{\Delta F_1}{k_{\rm B}T} \sim \Phi_{\rm C}\left(\frac{R_{\rm g}}{R_{\rm C}}\right)^{1 / \nu}.
\label{cage}
\end{equation}
In the limit $\Phi_{\rm C}\to 1$, Eq. \ref{cage} becomes the well known result for the free energy of confining a polymer in 
a slit-like cavity of size $R_{\rm C}$. It is important to remember that both Eq. \ref{combo} and Eq. \ref{cage} are valid only in the  limit  $R_{\rm C}/R_{\rm g}\to 0$. 

For macromolecular crowding in a cellular environment, the relevant regime is $R_{\rm C}\sim R_{\rm g}$. 
In this case, we have $\Delta F_1\sim k_{\rm B}T$ in Eq. \ref{combo} and
\begin{equation}
\frac{\Delta F}{k_{\rm B}T}\sim n
\label{Fn}
\end{equation}
in Eq. \ref{cage}.
Therefore, we anticipate that the free energy of the polymer will increase linearly with the number of crowders at the
polymer surface, $n$. Furthermore, for $R_{\rm C}\sim R_{\rm g}$, the volume excluded to the crowders by the polymer,
$V_{\rm exc}$, is not well approximated by $R_{\rm g}^3$ as was done in Eq. \ref{n1}. The actual value of $V_{\rm exc}$ will 
increase sharply with the crowder size $R_{\rm C}$, yielding a weaker dependence of $n$ on $R_{\rm C}$ than 
$R_{\rm C}^{-3}$ in Eq. \ref{n1}. 

\section*{Simulations}
In order to test the theoretical predictions,  we performed coarse-grained (CG) simulations of $\lambda_{6-85}$ repressor (see Methods for details) whose folding under infinite dilution conditions and {\it in vivo} have  been experimentally investigated \cite{Ghaemmaghami01NSB,Prigozhin11JACS,}.  We assume that the analysis presented above applies to the unfolded state of the 
protein, whereas the folded state is not significantly affected by the crowders. Then, Eq. \ref{Fn} determines the increase in the stability of
the protein due to crowding.  A relationship similar to Eq. \ref{Fn} can also be written for the increase in the melting 
temperature of the protein, $\Delta T_{\rm m}$, as long as the crowding effect is a relatively small perturbation. The use of  $C_{\alpha}-SCM$ is fully justified here because our theoretical arguments suggest that the predictions for $\Delta T_m(\Phi_C)$ are universal depending only globally on the characteristics of the unfolded state. In a number of applications, we have shown that  simulations based on $C_{\alpha}-SCM$ captures the folding reactions accurately \cite{Klimov00PNAS,OBrien08PNAS,Reddy12PNAS}.

The peak in the temperature dependence of the heat capacity (upper left corner of Fig. 2) is associated with the melting temperatures. The shift in the melting temperature, $\Delta T_m (\Phi_C)$, shown in green circles in Fig. 2, can be fit extremely well using Eq. (3) with $\alpha \approx$ 1.5 (black line in Fig. 2), which  is within the predicted bounds. The effective exponent $\nu_{eff} \approx$ 0.55. Our previous simulation results for effect of spherical crowding particles on the WW domain showed that $\Delta T_m(\Phi_C) \approx \Phi_C^{1.8}$ leading to $\nu_{eff} \approx$ 0.52. For both these systems the predicted bounds are satisfied further justifying the validity of the entropic stabilization mechanism. More importantly, theory and simulations in Fig. 2 show that $\Delta T_m (\Phi_C=0.15)$ is $\approx$3$^{\circ}$, which is very good agreement with experiments (experimental value for $\Delta T_m (\Phi_C=0.15)$ is  $\approx$4$^{\circ}$ \cite{Denos12FaradDisc} in which Ficoll was used as a crowding agent. We find the agreement particularly satisfying because we did not adjust any parameter to fit the experimental data.

The simulation data confirm that the melting temperature of $\lambda_{6-85}$ increases linearly with the number of 
crowders localized at the protein surface, $n$ (Fig. \ref{n_4panels}a). The increase in the melting temperature is approximately 
1 $^{\circ}$C per crowder. For given $\Phi_{\rm C}$ and $R_{\rm C}$, the value of $n$ was obtained by directly counting the number of
crowders whose centers were within distance $4R_{\rm C}/3$ from the protein surface. Each reported value represents an 
average over 10,000 protein conformations in the unfolded state at 105 $^{\circ}$C. We find that $n$ follows the power law,
\begin{equation}
n\propto\left(\frac{\Phi_{\rm C}}{R_{\rm C}}\right)^{1.43},
\label{F143}
\end{equation}
for all considered combinations of $\Phi_{\rm C}$ and $R_{\rm C}$ (\ref{n_4panels}b), and hence
$\Delta T_{\rm m}\propto\left(\frac{\Phi_{\rm C}}{R_{\rm C}}\right)^{\alpha}$, 
where $\alpha=1.43$.

In Fig. \ref{n_4panels}c, the dependence of $n$ on $\Phi_{\rm C}$ at $R_{\rm C}=24$ \r{A} (red symbols) is compared to the estimate 
$\rho_{\rm C}V_{\rm exc}$ (pink symbols), where $V_{\rm exc}$ was computed numerically for the same set of 10,000 
simulation snapshots that were used to determine $n$. Although the estimate $\rho_{\rm C}V_{\rm exc}$ is found to be 
fairly accurate, it specifies a linear dependence of $n$ on $\Phi_{\rm C}$ as opposed to the observed nonlinear 
dependence (\ref{n_4panels}c). These results indicate that inter-particle correlations in crowding suspensions promote 
clustering of crowders around the protein surface. 

Fig. \ref{n_4panels}d shows $n(R_{\rm C})$ at $\Phi_{\rm C}=0.25$, as obtained by direct counting (green symbols) or from the 
estimate $\rho_{\rm C}V_{\rm exc}$ (pink symbols). Both curves decay more slowly than $R_{\rm C}^{-3}$ specified in 
Eq. \ref{n1}, illustrating a large contribution of the crowder size $R_{\rm C}$ to the excluded volume $V_{\rm exc}$. 

We conclude that the value of $\alpha$ in Eq. \ref{Tm} is determined by crowder-protein and crowder-crowder correlations 
and as such it cannot be expressed in terms of the polymer scaling exponents alone although the precise value of $\alpha$ extracted from simulation satisfies the expected bounds. Thus, experiments can be analyzed using Eq. \ref{Tm} using a single adjustable parameter, the effective exponent $\alpha$. 

\section*{Comparison with experiments}
Waegle and Gai\cite{Gai11JCP} found empirically that the melting temperature of a 76-residue protein Ubiquitin(Ub) increases algebraically  as a function of $\Phi_C$ \cite{Cheung05PNAS} and is well  fit using Eq.(3). They used dextran of differing molecular weights and Ficoll 70 as crowding agents. The $\Phi_C$ dependent shift in the melting temperature was obtained from thermal melting curves inferred from Fourier transform infrared measurements. The measurements of $\Delta T_m(\Phi_C)$ as a function of $\Phi_C$ confirm the theoretical prediction with $\alpha$ ranging from $1.4 \le \alpha \le 2.1 \pm 0.1$. This implies that $0.5 \le \nu_{eff} \le 0.6$, which brackets the predicted theoretical bound. Interestingly, $\nu_{eff}$ =0.5 is obtained for high molecular particles Dextran40 and Dextran60. The finding that $\nu_{eff}$ = 0.5 implies that the crowding particles poises the solvent to be close to the $\Theta$-point.

It is more difficult to compare the dependence of $\Delta T_m(\Phi_C)$ on the crowder size because it is likely that the crowding particles considered in \cite{Gai11JCP} are not  spherical. Nevertheless, the observed weak dependence of $\Delta T_m(\Phi_C)$ on $R_C$  is qualitatively consistent with the theoretical predictions. It can be shown that the maximum effect of crowding  arises when $R_C \le R_g$. When $R_g \ge R_C$, the stabilities become independent of $R_C$ because in this limit the polypeptide chain behaves as though it is trapped between slits. In the case of Ub \cite{Gai11JCP}, the ratio $\frac{R_g}{R_C}$ is largest for Dextran 6 and is less than unity for higher molecular higher weight Dextran.  For Ficoll 70, $\nu_{eff} \approx$ 0.5 based on $\alpha \approx$ 2.0, which also suggests that $\frac{R_g}{R_C} < 1$.  Therefore, it is not surprising that there is only a weak dependence of $\Delta T_m(\Phi_C)$ on the size of the crowding particles. Our theory provides a quantitative explanation of the experiments on Ub.

\section*{Role of non-specific attractions}

A few experiments have reported that proteins can be destabilized (mildly) under  conditions that apparently mimic cellular conditions. Based on these findings it has been concluded that the predictions of the crowding theory, which only consider steric interactions, must be incorrect.   We first provide a theoretical argument showing that, even in the presence of non-specific attractive interactions between crowding particle and a polypeptide chain in a 
cellular environment, the macromolecular interactions are likely to be dominated by excluded volume effects. Let $u(r)$ be the interaction potential between residues 
in a protein (or RNA) and macromolecular crowders, such that $u(r)$ has an excluded volume part and an attractive part. The relative strength of the two 
contributions in $u(r)$ can be assessed using the second virial coefficient $B_2$,
\begin{equation}
B_2=\frac{1}{2}\int_{0}^{\infty}\left[1-\exp\left(-\frac{u(r)}{k_{\rm B}T}\right)\right],
\label{B2}
\end{equation}
where $T$ is the temperature and $k_{\rm B}$ is the Boltzmann constant. The second virial coefficient vanishes if the attractive interaction 
exactly balances the excluded volume contribution. To derive an analytical expression for $B_2$, we assume that $u(r)$ is a square-well potential comprising of both hard and soft interactions (Fig.~\ref{eps0_fig}),
\begin{eqnarray}
u(r)&=&\infty, r < D, \nonumber \\
u(r)&=&-\varepsilon, D \le r \le D+\sigma, \nonumber \\ 
u(r)&=&0, r > D + \sigma,
\label{POT}
\end{eqnarray}
where $D$ is the contact distance, corresponding to the universally relevant excluded volume, and $\sigma$ measures the range of the non-specific attractions. In the case of a protein
interacting with other crowding proteins, we estimate $D=R_{\rm a}+R_{\rm C}$ and $\sigma=2R_{\rm a}$, where $R_{\rm a}$ and $R_{\rm C}$ are the radii of the amino acid and 
crowder respectively. From Eqs.~(\ref{B2}) and (\ref{POT}) we find
\begin{equation}
\frac{3B_2}{2\pi D^3}=\left(1+\frac{\sigma}{D}\right)^3-\exp\left(\frac{\varepsilon}{k_{\rm B}T}\right)\left[\left(1+\frac{\sigma}{D}\right)^3-1\right].
\label{B2_2}
\end{equation}
The strength of non-specific attractions that is sufficient to neutralize the excluded volume effects, $\varepsilon_0$, follows by setting $B_2=0$ in Eq.~(\ref{B2_2}) yielding,
\begin{equation}
\frac{\varepsilon_0}{k_{\rm B}T}=-\ln\left[1- \left(1+\frac{\sigma}{D}\right)^{-3}\right].
\label{eps0}
\end{equation}
For any other form of the interaction potential $u(r)$ that is short-ranged, the attraction strength $\varepsilon_0$ will also depend only on the ratio of length scales $D$ and $\sigma$. 
The value of $\varepsilon_0$ increases sharply with decreasing $\sigma$ (Fig.~\ref{eps0_fig}b), indicating that strong attractive interactions are required for complete 
neutralization of the excluded volume effects when $\sigma/D<1$. We note that the condition $\sigma/D<1$ is satisfied for all realistic values of $R_{\rm a}$ and 
$R_{\rm C}$. 

Setting $R_{\rm C}=3R_{\rm a}$ to model a crowder protein of a minimum size, we find $\varepsilon_0/k_{\rm B}T\approx0.35$. This value is unrealistically large for any 
non-specific attractions that may be present between macromolecules. To illustrate this point, we observe that most biomolecules melt at temperatures that are only 
insignificantly higher (on the scale of $k_{\rm B}T$) than the physiological temperature. Assuming that the second virial coefficient of interactions between two 
amino acids vanishes at the melting temperature, we can use Eq.~(\ref{eps0}), with $D=\sigma$, to estimate the strength of non-specific attractions between 
amino acids. We find $\varepsilon_0/k_{\rm B}T\approx0.13$, which is of the order of the interaction strength between two carbon atoms, and approximately 3 times weaker 
than necessary to negate the crowder-amino acid excluded volume effect. Thus, for any realistic $R_{\rm C}$, $R_{\rm a}$ and $\varepsilon_0$ the value of $B_2$ is positive and likely
dominated by steric interactions. 

There are additional arguments which put an upper limit on the strength of non-specific attractions between macromolecules in the cell
and justify the positive values of $B_2$. First, in the case $B_2<0$, macromolecular crowders can adsorb onto an unfolded protein and interfere with its folding. (Such unwarranted interactions may be prevented by ATP-consuming chaperones, which are not germane to the experiments discussed below.) Furthermore, 
suspensions of macromolecular crowders with sufficiently negative $B_2$ and volume fractions representative of those in the cell (0.2--0.4) could form crystalline phases.
Clearly, these physical phenomena do not satisfy the conditions of non-specificity. We therefore conclude that, even if non-specific attractions are present between 
macromolecules in a cellular environment, their effect must be small compared to steric interactions.

{\bf Analyses of experiments - Numbers matter:} In light of the arguments above how can one understand the claims that {\it in vivo} environment or cell lysates destabilize proteins? This can only be done by examining the experimental results quantitatively using an analytic theory that captures one limiting case quantitatively, as we have done here.  It is useful to remind the readers that crowding theory (the one which examines only the consequences of steric interactions) predicts that the native state would be stabilized with respect to the $\Phi_C =0$ situation, resulting in negative  $\Delta \Delta G = \Delta G(\Phi_C) - \Delta G(0) < 0$ where $\Delta G$s are free energy differences between the folded and unfolded states. According to crowding theory, based on the concept of depletion interaction,  the magnitude of $\Delta \Delta G$ depends on the ratio $\frac{R_g(0)}{R_{\rm C}}$, and is negligible when $\frac{R_g(0)}{R_{\rm C}}$ is less than unity. 

A cell lysate presumably contains macromolecules of differing sizes (a polydisperse mixture), which makes it non-obvious on how one ought to choose $R_{\rm C}$. Our theory and related works~\cite{Minton05BJ}, rooted in the concept of depletion forces, predicts that the polypeptide chain in a polydisperse soup of macromolecules is most likely surrounded by large-sized crowders~\cite{Shaw91PRA} in order to maximize the total entropy of the system~\cite{Denesyuk11JACS,Kang15JACS}. In a cell lysate, we expect this be particles like ribosomes or other protein complexes with $R_{\rm C} \approx$ 10 nm, which is the approximate size of a ribosome. Thus, significant entropic stabilization is possible only if $R_g(0) \approx$ 10 nm.  With this simple description of the theory, we analyze experiments examining crowding effects on the stabilities of CI2 ($N_{aa} = 64; R_g(0) \approx 2.4nm$), $\lambda_{6-85}$ ($N_{aa} = 80; R_g(0) \approx 2.8nm$), CRABP ($N_{aa} = 136; R_g(0) \approx 3.8nm$), and VlSE ($N_{aa} = 341; R_g(0) \approx 6.6nm$), where $N_{aa}$ is the number of amino acids and the radius of gyration of the unfolded state in the absence of crowders is obtained using $R_g(0) \approx 0.2N_{aa}^{\nu}$nm with $\nu \approx 0.6$. Because the $R_g(0)$ values are not large compared to the size expected for significant stabilization, crowding theory would predict that $\Delta \Delta G$ to be negligible at most values of $\Phi_c$ of interest.  

The reported values of $\Delta \Delta G$s for these proteins in a cell lysates or cell-like environment are (in units of kcal/mol) -0.6$\pm$0.1 \cite{Sarkar13PNAS}, $\approx 0$, -0.2, and  -0.5 for CI2, $\lambda_{6-85}$ \cite{Tai16FEBSLett}, CRABP I (cellular retinoic acid-binding protein I) \cite{Ignatova04PNAS}, and VlsE \cite{Guzman14JMB} respectively. These values are small (less than $k_BT$). In three cases they suggest destabilization of proteins. Unless there are cases where the extent of destabilization is greater,  we interpret that these results are not inconsistent with crowding theory in contrast to the advertisement in the experimental papers. We believe that the onus is on the experimentalists to produce examples where the extent of destabilization exceeds $k_BT$. We hasten to add that destabilization of proteins is not ruled out if non-specific attraction is strong enough (see Eq. \ref{eps0}). However, the currently available examples are not the best ones to illustrate the extent of destabilization in cells. From these experiments one can infer that cell lysates do not significantly the stabilities of proteins, which can be accommodated using the crowding theory based on the idea of depletion interaction.

\section*{Conclusions}

If the volume excluded to the proteins by the crowders is the predominant interaction  then the complicated problem of protein stability, expressed using the dependence of the melting temperature on $\Phi_C$ is accurately predicted using Eq. (3), thus providing a firm theoretical framework. Surely, this is an  interesting limiting case. Remarkably, the exponent $\alpha$ is predicted to be universal and depends only on the properties of the unfolded state at $\Phi_C$ = 0. A few additional remarks are worth making. (1) The effects of excluded volume interactions, at a given $\Phi_C$, should give an upper bound to the enhancement in protein stability. Other favorable interactions, which alter the enthalpy of interactions, could diminish the stability of proteins, thus neutralizing the effect of steric interactions both on protein stability \cite{Ghaemmaghami01NSB} and association between proteins \cite{Rosen11JPCB}. (2) The theoretical predictions made here require that the polymer fluctuations in the unfolded state determine the dependence of $\Delta T_m(\Phi_C)$ on $\Phi_C$. Hence, ideas based on Scaled Particle Theory, which can capture some aspects of crowding, cannot be used to describe many features described here and elsewhere \cite{Denesyuk11JACS,Cheung05PNAS} even when non-specific attractions could be neglected. (iii) In analyzing experiments and simulations the exponent $\alpha$ should be chosen to fit the experimental data. The demonstration that for $\lambda_{6-85}$ the value of $\alpha$ cannot be tidily predicted from well-known polymer scaling exponents show that it should be considered to be a fit parameter. The bounds on $\alpha$ follow from polymer scaling exponents. (iv) In order to establish that realistic soup of cellular milieu (a multi component system) destabilize proteins two conditions must be met. First, the extent of destabilization must exceed $k_B T$. Second, under these conditions it must be demonstrated that the native state has not been altered. It may be the case that there are changes in the structures of the folded states and intermediates in cellular environment (macromolecules and various osmolytes) as atomic detailed simulations \cite{Feig17JPCB} seem to suggest.  

{\bf Polydisperse Effects:} Perhaps, the most interesting aspect of the theory, based on the entropic stabilization of the folded state, is its utility in providing an alternate explanation for the apparent near universal destabilization of proteins in cell-like environments. It should be emphasized that theory {\it does not preclude the possibility that crowders destabilize proteins}. Two scenarios could be envisaged. ({\bf I}) The popular explanation is that soft interactions between crowders and proteins  \cite{Cheung06JMB,Drishti19ChemRev,Sarkar13PNAS,Sapir15CurrOpinColl,Sapir14JPCL,Danielsson15PNAS}) result in negation of the stabilizing effect due to volume exclusion. For this to occur the interaction strength has to be substantial (see Eq.\ref{eps0} and \cite{Rosen11JPCB} for an estimate for a protein complex).  We had previously shown that under these conditions the native state may not be stable and other states, including the possibility that the protein would weakly adsorb onto the crowders \cite{Cheung06JMB}, are populated with higher probability. Under these conditions, which apparently is realized in atomic simulations of folding in the presence of crowders, comparing of the free energy changes in the crowded milieu is not meaningful. ({\bf II}) It is also possible that crowders, interacting with the protein solely through excluded volume interactions, largely affect only the unfolded state  without affecting the folded state. In this scenario, crowders would render the unfolded state compact, as predicted theoretically \cite{Edwards88JCP,Thirum88PRA}, facilitating intra peptide attraction. This would stabilize the unfolded state enthalpically increase the barrier to folding. As a consequence, the stability would decrease and the folding time is predicted to be slower. This scenario explains the modest changes in stability and folding time for VLSE and PGK in cells relative to {\it in vitro} (see Table 1 in \cite{Tai16FEBSLett}).  ({\bf III}) The modest destabilization observed in several experiments is usually rationalized in terms of weak interactions, which of course is a possibility that cannot be ruled out. Based on the simulations on DNA flexibility \cite{Kang15JACS} and theoretical considerations \cite{Kang15PRL}, we believe that the concept of depletion forces when extended to polydisperse crowding agents offer an alternative explanation. Consider a mixture of crowding agents with different sizes and shapes, which might be a better mimic of the cytoplasm. The AO picture predicts that maximization of entropy is realized if the protein is surrounded by the largest crowding particles. If this were the case then the stability would not be altered significantly \cite{Denesyuk11JACS}, which would not be inconsistent with experimental data analyzed here. A potential {\it in vitro} experiment which could shed light on this issue is to use Ficoll and Dextran of various sizes (with some crowding particles exceeding the radius of gyration of the unfolded states), and assess the changes in the melting temperatures.

\subsection{Methods}

{\bf Model:} The polypeptide chain is simulated using a coarse-grained model, in which each amino acid is replaced by two spherical beads, 
representing a $C_{\alpha}$ atom and a side chain ($SC$) \cite{Klimov00PNAS}. The use of  $C_{\alpha}-SC$ model is fully justified here because our theory suggests that the predictions for $\Delta T_m(\Phi_C)$ and the associated stabilities  are universal depending globally only on the characteristics of the unfolded state. The energy function in the $C_{\alpha}-SC$ coarse-grained model, $U_{\rm CG}$, has the following four 
components,
\begin{equation}
U_{\rm CG}=U_{\rm CC}+U_{\rm SC}+U_{\rm EV}+U_{\rm NAT},
\label{CG}
\end{equation}
corresponding to $C_{\alpha}-C_{\alpha}$ and $C_{\alpha}-SC$ bond length constraints, excluded volume repulsions and native interactions. Bond lengths
are constrained by harmonic potentials, $U_{\rm CC}(\rho)=k(\rho-\rho_{\rm CC})^2$ and $U_{\rm SC}(\rho)=k(\rho-\rho_{\rm SC})^2$, 
where $k=30$ kcal$\,$mol$^{-1}$\r{A}$^{-2}$. The equilibrium distance between two $C_{\alpha}$ atoms, $\rho_{\rm CC}$, is defined individually for each bond as the corresponding 
$C_{\alpha}$-$C_{\alpha}$ distance in the protein crystal structure (PDB code 1LMB). Similarly, $\rho_{\rm SC}$ is the distance between each 
residue's side chain center of mass and $C_{\alpha}$ atom in the PDB structure.

We model the excluded volume interactions between two coarse-grained beads $i$ and $j$ separated by distance $r$ using a generalized Weeks-Chandler-Andersen (WCA) potential,
\begin{eqnarray}
U_{\rm EV}(r)&=&\varepsilon_{\rm EV}\frac{{\rm min} (D_i, D_j)}{D_{\rm C}}
\left[\left(\frac {D_{\rm C}}{r + D_{\rm C} - D_{ij}}\right)^{12} 
- 2\left(\frac {D_{\rm C}}{r + D_{\rm C} - D_{ij}}\right)^{6} + 1\right],\ r\le D_{ij}, \nonumber \\
U_{\rm EV}(r)&=&0,\ r > D_{ij}, 
\label{UEV}
\end{eqnarray}
where ${\rm min} (D_i, D_j)$ is the smaller of the bead diameters $D_i$ and $D_j$, $D_{ij}=0.5(D_i+D_j)$ and $\varepsilon_{\rm EV} = 1$ kcal/mol. 
The diameter of a bead representing a $C_{\alpha}$-atom is $D_{\rm C}=3.8$\r{A}. For each residue, the value of $D_{\rm S}$ is given by 
$V_{\rm S} = \pi D_{\rm S}^3/6$, where $V_{\rm S}$ is the van der Waals volume of the side chain computed from the PDB coordinates of its
individual atoms using the AMBER94 atomistic van der Waals radii. Thus, side chains of the same kind can have somewhat different $D_{\rm S}$ 
depending on their configuration in the PDB structure. Interactions of crowders with the protein residues and other crowders are modeled using the
same potential as in Eq.~(\ref{UEV}). We chose the generalized form of the WCA potential because it is well suited to model excluded volume between two 
particles with very different diameters, such as a $C_{\alpha}$-atom and a macromolecular crowder. The ratio ${\rm min} (D_i, D_j)/D_{\rm C}$ in Eq.~(\ref{UEV}) 
rescales the interaction strength $\varepsilon_{\rm EV}$ in proportion to the surface contact area between particles $i$ and $j$.

Native interactions between protein coarse-grained beads are modeled by a Lennard-Jones potential, 
\begin{eqnarray}
U_{\rm NAT}(r)=\varepsilon_{\rm NAT}\left[\left(\frac {r_0}{r}\right)^{12} - 2\left(\frac {r_0}{r}\right)^{6}\right],
\label{UNAT}
\end{eqnarray}
where $r_0$ is the corresponding interbead distance in the coarse-grained PDB structure. The native interactions are defined only for those pairs of beads 
for which $r_0<8$ \r{A}. The value of $\varepsilon_{\rm NAT}$ is the same for all native interactions, $\varepsilon_{\rm NAT}=0.43$ kcal/mol, and has been adjusted
so that the experimental melting temperature of $\lambda_{6-85}$ is reproduced in simulation ($T_{\rm}=56^\circ$C).

{\bf Equations of motion:} The protein and crowder dynamics are simulated by solving the Langevin equation, which for bead $i$ is 
$m_i\ddot{\mathbf{r}}_i=-\gamma_i\dot{\mathbf{r}}_i+\mathbf{F}_i+\mathbf{f}_i$, where $m_i$ is the bead mass, 
$\gamma_i$ is the drag coefficient, $\mathbf{F}_i$ is the conservative force, and $\mathbf{f}_i$ is the Gaussian random 
force, $\left<\mathbf{f}_i(t)\mathbf{f}_j(t^{\prime})\right>=6k_{\rm  B}T\gamma_i\delta_{ij}\delta(t-t^{\prime})$. The mass of $C_{\alpha}$ bead is
the molecular weight of a carbon atom and the mass of a $SC$ bead  is the total molecular weight of the corresponding side chain. We use the mass
of 8.6 kDa for a crowder with diameter 24 \r{A}, which is consistent with typical protein densities. The mass of any other crowder $i$ with diameter 
$D_i$ is estimated as $m_i=8.6(D_i/24)^3$ kDa. The drag coefficient $\gamma_i$ is given by the Stokes formula, $\gamma_i=3\pi \eta D_i$, where $\eta$ 
is the viscosity of the medium. To enhance conformational sampling~\cite{Honeycutt92Biopolymers}, we take 
$\eta=10^{-5}$Pa$\cdot$s, which equals approximately 1\% of the viscosity of water. The Langevin equation is integrated using the leap-frog algorithm 
with a time step $\Delta t=10$ fs.

{\bf Acknowledgments:} This work was done while the authors were in the Institute for Physical Sciences and Technology at the University of Maryland. We appreciate the useful comments from Martin Grubele and Ben Schuler.  We are grateful to the National Science Foundation (CHE 19-00093) and the Collie-Welch Chair (F-0019) for supporting this research.

\newpage


\newpage
\begin{center}
\textbf{\large{Figure Captions}}
\end{center}

{\bf Figure~1} Snapshot of the encapsulated $\lambda$-repressor  from coarse-grained SOP simulations in the presence of crowding particles shown in dark blue. The protein is localized in a roughly spherical region with diameter $D$. This picture provides a physical basis for approximate mapping between crowding and confinement. 

{\bf Figure~2} Dependence of the shift in the melting temperature ($\Delta T_m(\Phi_C)$ for $\lambda$-repressor as a function of $\Phi_C$. The green circles are obtained from simulations and the black line is a theoretical fit using Eq. (3) with $\alpha$ = 1.5. The structure on the right is a ribbon representation of the native state of the protein. The melting temperatures are associated with the peaks in the heat capacity such as the ones shown for $\Phi_C$ = 0 (black curve) and $\Phi_C$ = 0.3 (dashed red curve).  

{\bf Figure~3} The increase in the melting temperature, $\Delta T_{\rm m}$, and the number of crowders localized at 
the protein surface, $n$, obtained in simulations of $\lambda$ repressor. The parameter $n$ is defined as the number 
of crowders positioned within distance $4R_{\rm C}/3$ from the protein surface. Red squares show simulation results for 
$R_{\rm C}=24$ \r{A} and $\Phi_{\rm C}=0.05$, 0.1, 0.15, 0.2, 0.25, 0.3. Green circles are for $\Phi_{\rm C}=0.25$ and 
$R_{\rm C}=12$, 20.5, 24, 41, 48, 96 \r{A}. Pink symbols in (c) and (d) show $n$ as estimated by $\rho_{\rm C}V_{\rm exc}$,
where $V_{\rm exc}$ is the protein-crowder excluded volume computed numerically from simulation snapshots. The solid lines 
correspond to functions (a) $y=1.08 x$, (b) $y=3452 x^{1.43}$, (c) $y=36.5 x^{1.43}$, and (d) $y=479 x^{-1.43}$.

{\bf Figure~4} (a) Square-well potential, $u(r)$, used in the calculation of the second virial coefficient, $B_2$. (b) $\varepsilon_0$ 
 as a function of $\sigma/D$. By definition $B_2$ vanishes at $\varepsilon=\varepsilon_0$.

\newpage
\begin{figure}[ht]
\includegraphics[width=4.00in]{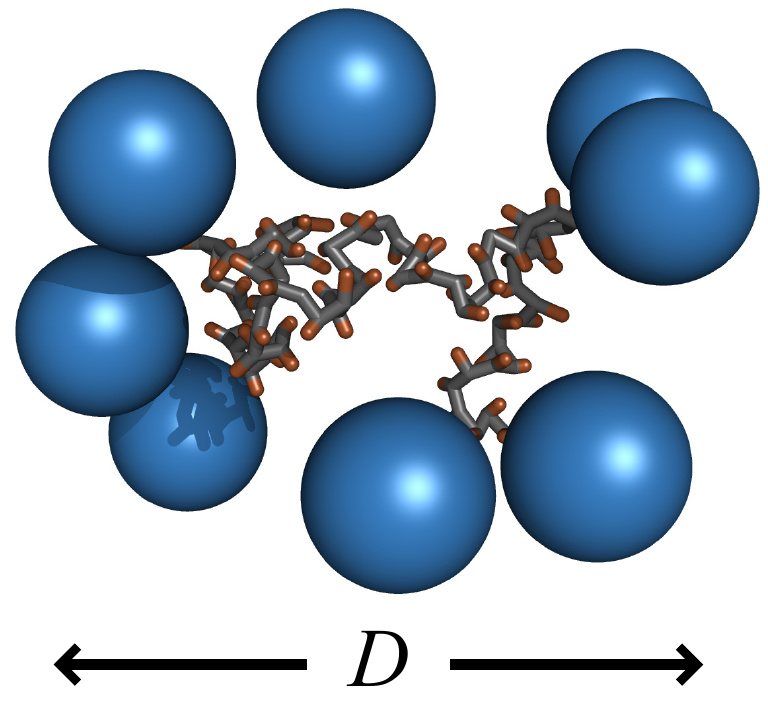}
\caption{\label{figure1}}
\end{figure}

\newpage
\begin{figure}[ht]
\includegraphics[width=4.00in]{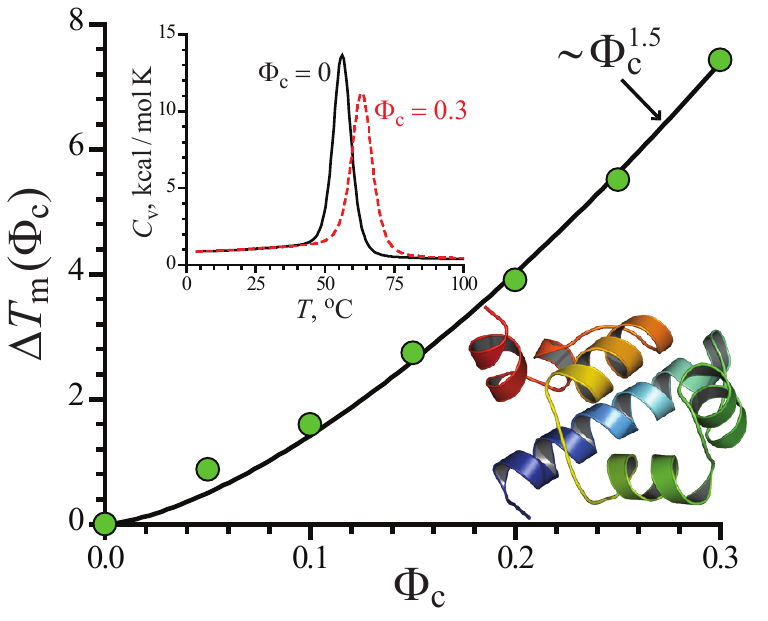}
\caption{\label{figure2}}
\end{figure}

\newpage
\begin{figure}[ht]
\includegraphics[width=4.00in]{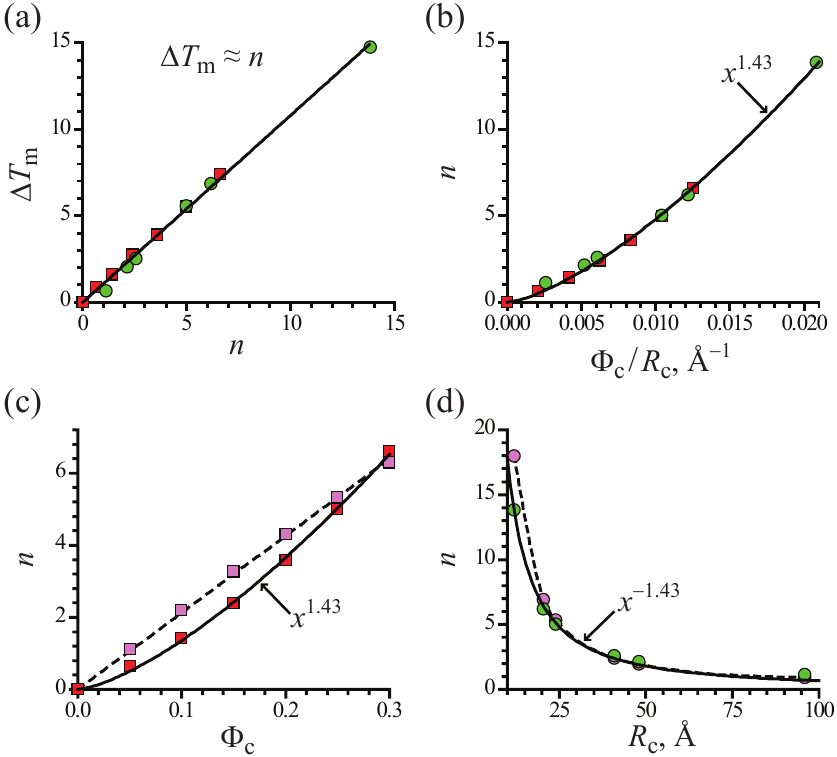}
\caption{\label{n_4panels}}
\end{figure}

\begin{figure}
\includegraphics[]{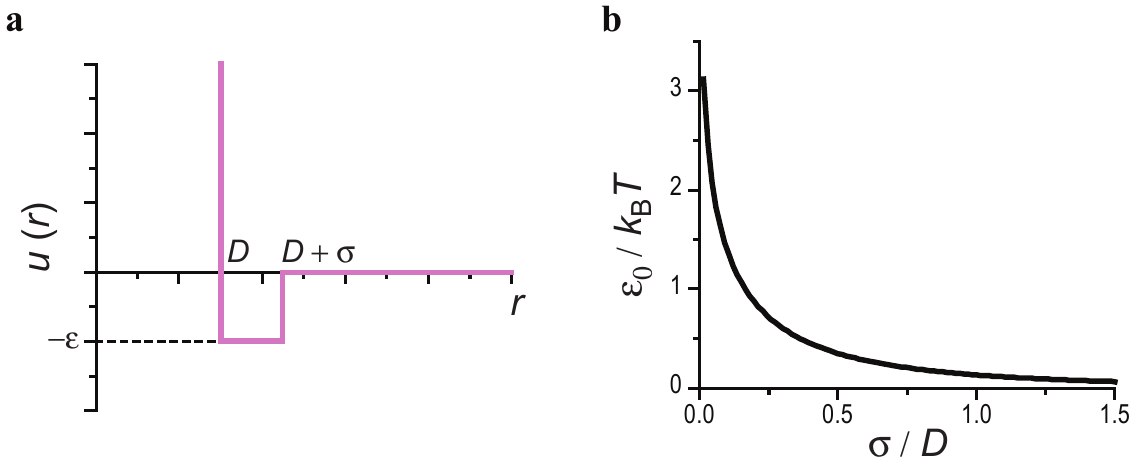}
\caption{
\label{eps0_fig}}
\end{figure}

\end{document}